# Experimental verification of a plasmonic hook in a dielectric Janus particle

*Igor V. Minin[1,2], Oleg V. Minin[1,2], Igor A. Glinskiy[3], Rustam A. Khabibullin[4], Radu Malureanu[5], Dmitry I. Yakubovsky[6], Valentyn S. Volkov[6], Dmitry S. Ponomarev[4,6]*

[1]National Research Tomsk Polytechnic University, Tomsk 634050, Russia

[2]National Research Tomsk State University, Tomsk 634050, Russia

[3]Prokhorov General Physics Institute of the Russian Academy of Sciences, Moscow 119991, Russia

[4]Institute of Ultra High Frequency Semiconductor Electronics RAS, Moscow 117105, Russia

[5]Department of Photonics Engineering, Technical University of Denmark, Ørsteds plads, bldg. 345V, DK-2800, Kgs. Lyngby, Denmark

[6]Center for Photonics and 2D Materials, Moscow Institute of Physics and Technology (MIPT), Dolgoprudny, 141700, Russia

KEYWORDS

surface plasmon-polariton, plasmonic beams, Airy-beams, plasmonic hook, scanning near-field microscopy, plasmonic lens, finite element method, subwavelength imaging

ABSTRACT

We report on a very first experimental observation of a new curved plasmonic beam, a plasmonic hook (PH), for surface plasmon-polariton (SPP) waves. We demonstrate that the SPP PH effect could be realized within a trapezoidal dielectric microstructure with the relative ease of fabrication compared to the generation of the Airy SPP waves which require complex techniques to compensate the wave vectors mismatch. Using numerical analysis, we predicted the PH phenomenon, optimized the microstructure and then confirmed the SPP PH existence via amplitude and phase-resolved scattering scanning near-field optical microscopy at the telecom wavelengths of 1530–1570 nm. The experimental results are in reasonable agreement with our predictions. Importantly, the SPP PH demonstrates the smallest radius of curvature ever recorded for SPP waves compared to that for the Airy-family plasmonic beams providing many useful applications spreading from nanoparticles manipulation toward bio-sensing within a nanoscale.

1. **Introduction.**

Plasmonic structures are now being intensively studied due to the ability to squeeze light into objects with sub-wavelength size thanks to the nature of surface plasmon-polaritons (SPPs)[1]. The SPPs refer to a strongly localized collection motion of free electrons when being coupled to photons at a metal-dielectric interface[2]. They are essentially two-dimensional (2D) waves whose field components decay exponentially with the distance from the surface. Since these waves tightly cling to the surface, this allows implementing them for biosensing, micromanipulations[3], and molecules diagnosis[4].

The discovery of curved accelerating and self-bending Airy-like SPP beams has driven new opportunities for plasmonics, high-resolution imaging, biophotonics, and optical data storage. Among them are zeroth-order Bessel[5] and non-diffracting Bessel SPP beams,[6] as well as rapidly accelerating Mathieu and Weber surface plasmon beams[7]. For a long time, the Airy-like plasmons have been the only plasmonic waves that are self-healing and nonspreading (diffraction-free) and thus are unaffected by surface roughness and any structural imperfections [2]. Moreover, the Airy beams are used to be non-diffracting wave packets in one-dimensional (1D) planar systems[8] compared to traditional 2D non-diffracting beams[9].

Previously, we theoretically predicted a new curved plasmonic wave, a plasmonic hook (PH), that could be realized when the in-plane SPP wave is passing through the asymmetric dielectric Janus microstructure[10]. The curved shape of the PH arises from constructive interference between the incident, scattered, and diffracted fields near the dielectric microstructure due to the near-field interactions at the shadow boundary of the structure. It was shown that the PH propagates along a wavelength-scaled curved trajectory with a radius less than the SPP wavelength compared to the Airy-family SPP beams[11,12,13], representing the smallest radius of curvature ever recorded for an SPP beam that can exist despite the strong energy dissipation arising from large Ohmic losses in noble metals. Interestingly, the curved shape of the hook may appear even in free space and requires no wave-guiding structures or any external potential[14].

In 2019, the photonic hook was experimentally verified in free space by I.V. Minin et al[15]. In this letter, we report on the very first experimental realization of the PH phenomenon for the SPP waves. We show that the PH could be generated using a trapezoidal dielectric microstructure, combines prism refraction with cube diffraction, with the relative ease of fabrication compared to the generation of Airy surface plasmons which require complex techniques to compensate a wave vector mismatch between an SPP and a free-space wave[16]. The SPPs are excited with the telecom wavelengths $\lambda_0$ = 1530–1570 nm using amplitude and phase-resolved scattering scanning near-field optical microscope. The experimental results are well-agreed with our theoretical predictions.

Interestingly, the SPP PH demonstrates the smallest radius of curvature ever recorded for SPP waves compared to that for the Airy-family plasmonic beams. This feature is of importance for various applications spreading from light manipulation within a nanoscale toward microscopy, biosensing and on-chip photonic devices, as well.

2. **Simulations.**

To demonstrate the SPP PH effect, we use a trapezoidal microstructure comprising a wedge combined to a dielectric cube which is placed on a 100 nm thick gold film. The AR-P 6200 e-beam resist from Allresist GmbH was selected as a cube and wedge material platform (the dependence of its refractive index on the telecommunication wavelength is given in Supplementary data). We carry out a numerical analysis of the PH formation using a full-wave 3D simulation (a finite element method) by COMSOL Multiphysics.

In the simulation, the SPP wave is passing through the microstructure when being excited by an optical illumination at the telecom wavelength $\lambda_0$ = 1530 nm featuring a maximum electric field of $E$ = 1 Vm$^{-1}$ at the dielectric–metal interface (see **Fig. 1**). The cube has the lateral dimensions $l_x = l_z = w$ = 5085 nm, and the wedge' angle is equal to $\theta$ = 27°.

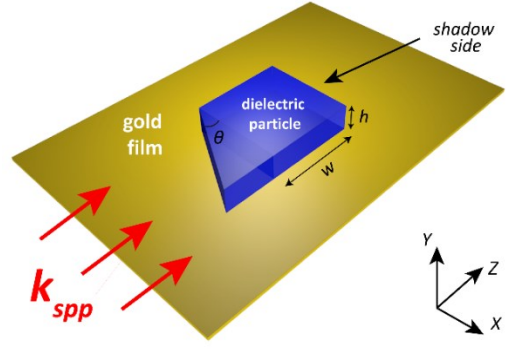

**Fig. 1.** Schematics of the SPP PH generation from the shadow side of Janus microstructure comprising a dielectric cube with a broken symmetry which is placed on a gold film. The SPP wave with a wave vector $k_{spp}$ is incident on a wedge's surface.

In the simulation, we use perfectly-matched layers' conditions and a non-uniform mesh featuring a minimum cell size of $1/5\lambda_0$ at a dielectric-metal interface. All the field distributions shown below are normalized to its maximum values. The simulation details can be found in [10,17]. We use a Drude–Lorentz dispersion model with relative permittivity of gold $\varepsilon_m = -114.47 + 8.51i$ at $\lambda_0 = 1530$ nm [18]. The SPP wavelength under the considered conditions is equal to $\lambda_{spp} = 0.978\lambda_0 = 1497$ nm (in other words, about 2% less than the wavelength in free space) [17].

**Figure 2** represents the SPP field intensity distributions $|E|^2$ within the $xz$-plane in the microstructure with different cube heights $h =$ 150, 250 and 350 nm.

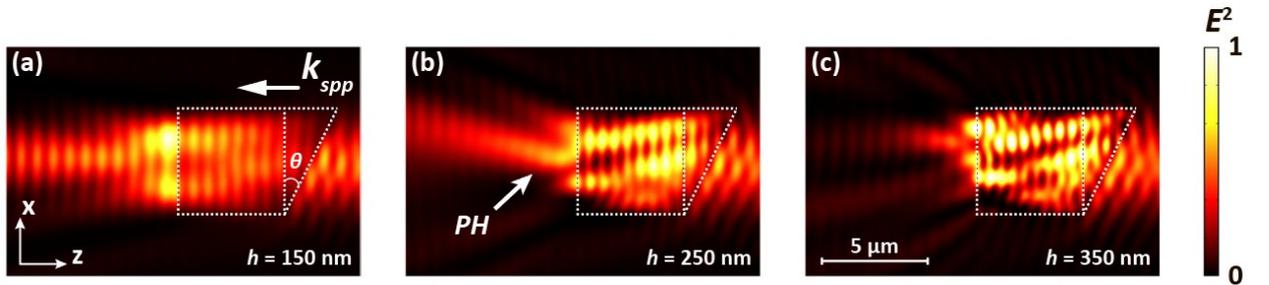

**Fig. 2.** SPP field intensity $|E|^2$ distributions in the microstructure vs. the cube height. The SPP wave with a wave vector $k_{spp}$ is incident on the microstructure from right to left and is excited with the telecom wavelength $\lambda_0 = 1530$ nm. The color bar demonstrates the intensity profile of the SPP PH propagation.

As seen in the figure 2, the formation of the hook is severely dependent on the cube height. For instance, at the low height, we observe a plasmonic jet [17], while the further increase in $h$ changes the distribution of the SPP field at the shadow side of the microstructure, providing a curved profile of the plasmonic wave. According to the simulation results, the maximum curvature of the SPP wave is observed at $h = 250$ nm which we further used in the fabrication of the microstructure.

### 3. Fabrication of the microstructure and experimental verification of the PH effect.

The microstructure was fabricated using electron-beam lithography. In the first instance, the plasmonic waveguide and the diffraction grating were exposed. Then, 100 nm of Au was deposited and patterned by a lift-off process using a positive photoresist. A second, aligned electron beam exposure was made to define the cube structure. Thanks to the precision of the electron beam technique, the placement of the cube was within +/- 20 nm from the desired position and the dimensions of the grating and the cube were within +/- 10 nm from the desired ones. These tolerances are far better than needed in this case. The nominal dimensions of the waveguide are 50×10 μm, the grating having a period of 1540 nm and a 50% filling fraction while the cube dimensions are $h = 220$ nm and width of $w = 5058$ nm. The cube is placed at 5 μm from the last slit in the grating.

The amplitude and phase-resolved scattering scanning near-field optical microscopy measurements were performed using the NeaSNOM from Neaspec GmbH (s-SNOM). The s-SNOM is working as an atomic force microscope in a taping mode with a sharp metal-coated silicon tip, as the near-field probe, oscillating at the resonance frequency of $\Omega \approx 280$ kHz with amplitude ~ 55 nm. To direct the SPP wave into the dielectric cube, the plasmonic grating coupler was illuminated from below by a linearly polarized light at a normal angle to the sample surface (transmission configuration). A telecom-wavelength semiconductor laser was used as a light source for near-field imaging. In this s-SNOM arrangement while mapping of near-field and topography across the scan area of 30×10 μm$^2$, an illumination system remains aligned on the diffraction due to its synchronization with the sample moving. The tip-scattered light is collected by a top parabolic mirror and goes to the detector. To provide clear imaging of near-field distribution most of the optical background was suppressed by demodulation of the detected signal at high-order harmonic frequency nΩ (*n* = 2, 3, 4) and by using interferometric pseudoheterodyne detection scheme with a modulated reference beam. In our case, the signal of demodulation at the third harmonic (3Ω) was taken into consideration, which was enough for background-free near-field analysis. **Figure 3** shows the experimental results for $\lambda_0$ = 1530 nm. As follows from (a), the plasmonic beam begins deviating starting from the inflection point and resulting in the PH effect. The panels (b) and (c) illustrate cross-sections of the SPP field intensity $|E|^2$ distributions near the shadow side of the microstructure in a focal spot (*x*-direction) and along the PH propagation axis (*z*-direction), respectively.

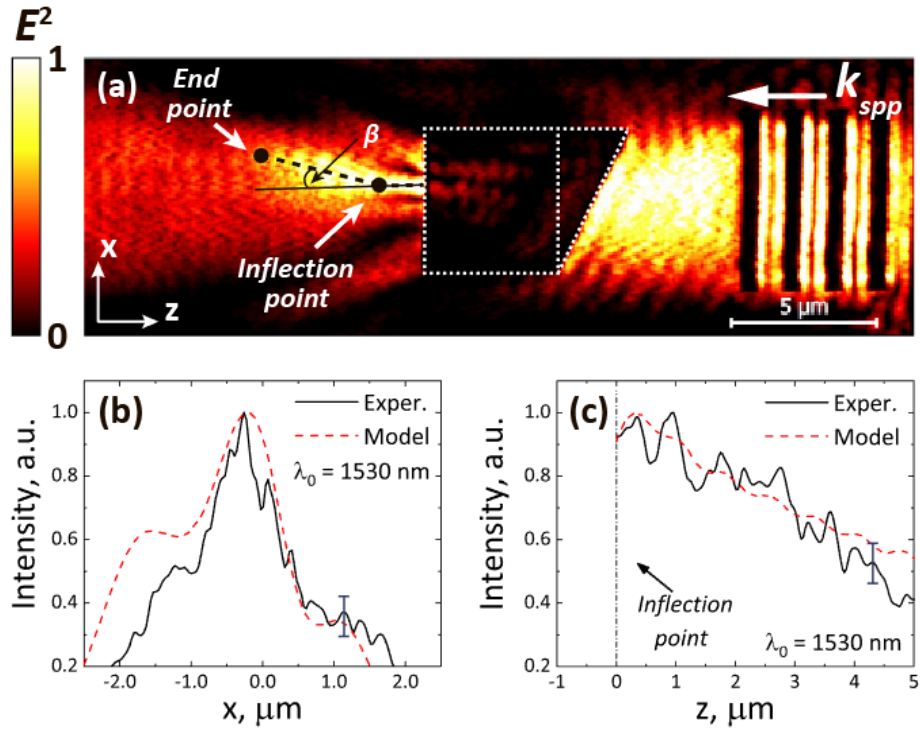

**Fig. 3.** SNOM images illustrating formation of the PH and high-intensity spots (a); cross-sections of the SPP field intensity $|E|^2$ distributions near the shadow side of the microstructure in a focal spot in the *x*-direction (b) and along the hook propagation axis in the *z*-direction (c). The SPP wave with a wave vector $k_{spp}$ is excited with the telecom wavelength $\lambda_0$ = 1530 nm. The SPP wave is incident on the microstructure from right to left (a). The color bar demonstrates the intensity profile of the SPP PH propagation.

According to [10,15], we define the curvature of the SPP PH as an angle of beam bending (β) as shown in figure 3a. From the experiment, it is followed that the curvature of the SPP PH is β = 16° for $\lambda_0$ = 1530 nm (in the simulation, we obtain β = 27° for $\lambda_0$ = 1530 nm). Interestingly, as seen in (a), the beam starts bending at ~ 1 μm distance from the shadow side of the microstructure (in other words, at the distance close or slightly above the wavelength of optical excitation) demonstrating the smallest radius of curvature ever recorded for SPP waves compared to that for the Airy-family plasmonic beams.

To demonstrate the tunability of our approach, we used two more optical excitation wavelengths $\lambda_0$ = 1550 nm and 1570 nm to excite the SPP wave and study the PH phenomenon using the s-SNOM. For convenience, the derived simulation and experimental data are summarized in **Table 1**. One can notice a reasonable agreement between the results. As follows from the table, a magnitude of full width at half maximum (FWHM), as well as the propagation length of the PH (*L*), defined as the length starting from the inflection point (the point when the plasmonic beam begins deviating from the propagation axis, see figure3a) to the end point (when the SPP beam shows an exponential decrease in its intensity, see figure3a), are increasing with an increase of the wavelength reaching their maximum experimental values FWHM = 1.342 μm (or 0.85$\lambda_0$) and *L* = 8.03 μm (or 5.11$\lambda_0$), respectively.

**Table 1**. Comparison between simulation and experiment results.

| Optical excitation wavelength $\lambda_0$, nm | Simulation | | Experiment | |
|---|---|---|---|---|
| | FWHM, μm | PH propagation length, *L*, μm | FWHM, μm | PH propagation length, *L*, μm |
| 1530 | 1.177 | 7.656 | 1.083 | 5.73 |
| 1550 | 1.213 | 7.838 | 1.166 | 6.02 |
| 1570 | 1.342 | 8.268 | 1.225 | 8.03 |

**Conclusion.**

In summary, we have demonstrated the very first experimental observation of a new curved sub-wavelength plasmonic beam, a plasmonic hook, that could be used to focus surface plasmon-polaritons (SPPs) through a trapezoidal dielectric microstructure which is placed on a thin metal film. By means of numerical simulations, we predicted the PH phenomenon and then confirmed its existence using amplitude and phase-resolved scattering scanning near-field optical microscopy within the telecom wavelengths $\lambda_0$ = 1530–1570 nm. The experimental results are in reasonable agreement with the simulation data.

The PH demonstrated the smallest radius of curvature ever recorded for SPP waves compared to that for the Airy-family plasmonic beams providing many useful applications like nanoparticles manipulation and bio-sensing within a nanoscale. Importantly to notice that such plasmonic beams could be used to increase photocarrier confinement in photoconductive devices operating at terahertz frequencies[19]. Nevertheless, this study lies beyond the scope of the present paper.

**Acknowledgments.**

I.A.G and D.S.P acknowledge the financial support of the Russian Scientific Foundation (Project No.18-79-10195), I.V.M. and O.V.M were partially supported by the Russian Foundation for Basic Research (Grant No. 20-57-S52001). R.M. acknowledges the support from the National Centre for Nano Fabrication and Characterization (DTU Nanolab) for the fabrication of the microstructures. The work was partially carried out within the framework of the Tomsk Polytechnic and Tomsk State Universities Competitiveness Enhancement Programs.

**Contributions.**
I.V.M. and O.V.M initiated the idea, supervised a project and proposed series of the research in this field, I.A.G and D.S.P made a simulations, D.S.P. wrote a draft, I.V.M., O.V.M and D.S.P. edit a draft, R.M., D.I.Y and V.S.V. made an experiment. All authors discussed results.

**Disclosures.**
The authors declare that there are no conflicts of interest related to this article.

**Supplementary data.**

In this section, we present some supplementary data.

**Figure S1** demonstrates the experimental dependence of the refractive index *n* on excitation wavelength $\lambda_0$ for the AR-P 6200 e-beam resist from Allresist GmbH used as the cube's material.

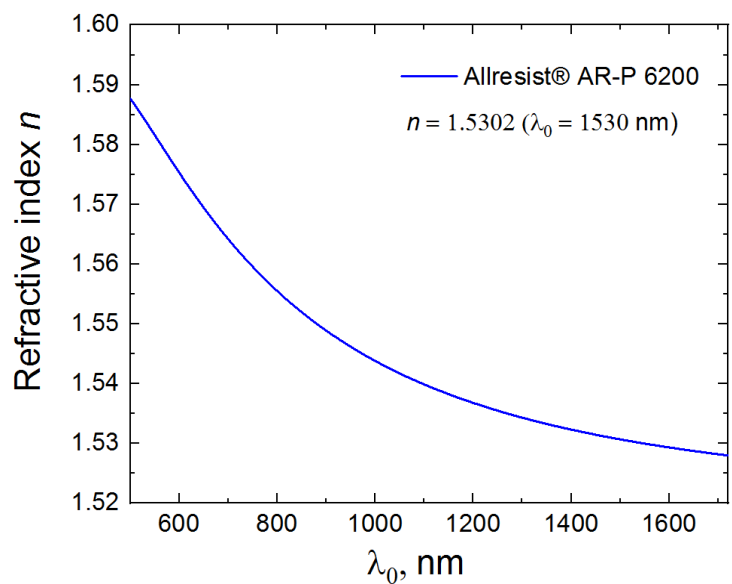

**Fig. S1.** Experimental dependency of the refractive index *n* on excitation wavelength $\lambda_0$ for the AR-P 6200 e-beam resist from Allresist GmbH used as the cube's material.